\documentclass[11pt]{article}
\usepackage{setspace}
\usepackage{amssymb}
\usepackage{amsmath}
\usepackage{amsfonts}

\def\be{\begin{equation}}

\def\ee{\end{equation}}

\def\bea{\begin{eqnarray}}

\def\eea{\end{eqnarray}}

\newcommand\eps{\epsilon}

\makeatletter
\def\blfootnote{\xdef\@thefnmark{}\@footnotetext}
\makeatother

\begin{document}

\singlespace

\begin{flushright} BRX TH-6308 \\
CALT-TH 2016-026
\end{flushright}

\vspace*{.3in}

\begin{center}

{\Large\bf One-loop gravity divergences in $D>4$ cannot all be removed}

{\large S.\ Deser}

{\it 
Walter Burke Institute for Theoretical Physics, \\
California Institute of Technology, Pasadena, CA 91125; \\
Physics Department,  Brandeis University, Waltham, MA 02454 \\
{\tt deser@brandeis.edu}
}
\end{center}

\begin{abstract}
Unlike the one-loop QG divergences in $D=4$, which can all be transformed away, those in arbitrary higher (even) $D$ cannot.  
\end{abstract}
 
\section{$D=4$}
Explicit loop calculations in quantum gravity (QG) originated over four decades ago with `tHooft and Veltman's [1] one-loop results in 4D. They noted that, even without calculation, the one-loop, hence quadratic in curvature, leading correction of the source-free theory could always be removed: First, the Riemann-squared term can be turned into a sum of Ricci- and scalar- curvature-squared ones using the pure divergence nature of the 4D Gauss-Bonnet (GB) invariant,
\begin{equation}
\int d^4 x \eps^{\alpha\beta\gamma\delta}  \, \eps^{\rho\sigma\mu\nu} \, R_{\alpha\beta\rho\sigma} \, R_{\gamma\delta\mu\nu} \sim \int d^4 x [R_{\mu\nu\rho\sigma} \, R^{\mu\nu\rho\sigma} - 4 \, R_{\mu\nu} \, R^{\mu\nu} + R^2 ] \sim 0.
\end{equation}
Second, the remaining, $\int R_{\mu\nu}^2$ and $\int R^2$, terms can then be relegated to higher order by the field redefinition
\begin{equation}
\delta g_{\mu\nu} =a(g, R) \, R_{\mu\nu} +b(g,R)\,  g_{\mu\nu} R,       
\end{equation}
one that is clearly available in all $D$ to remove monomials involving at least one Ricci tensor or scalar. [It is essential that the quadratic corrections are, by definition, NOT to be taken as part of the ``kinetic" term, which remains $R$, and of course cannot be removed.] 

The development of string theory and its corrections to QG in $D=10$ has made the study of higher $D$ relevant (see [2] for a recent example). We note here, in the same no-calculation spirit, that 
the no-correction result breaks down in $D=2n > 4$, because there is now more than one independent monomial in $(R_{\mu\nu\alpha\beta} )^n$, while there is still only one, GB ($D>4$), available to remove them. Even-parity counterterms (and GB) cannot exist at all in odd dimensions, since they would have to have an odd number of derivatives, hence of indices: Power counting in GR trivially establishes that all one-loop counterterms are of $D$-derivative order because vertices/propagators are universally of $+2$/$-2$ derivative order and cancel each other when adding a new vertex, so that the cutoff dependence is just that of the integration dimension, $\int d^D p$, for any number of external gravitons.  

\section{$D>4$}
The relevant GB invariant in any even, $D=2n$, (1) simply generalizes to 
\begin{equation}
GB(D=2n) = \int d^{2n} x \eps^{\ldots} \, \eps^{\ldots} (R_{\ldots}^1 \, \, \, \, R_{\ldots}^n) \sim 0,
\end{equation}
while of course (2) still removes all terms involving at least one Ricci or curvature scalar in any $D$. 
At the referee's request, here is the reason there is only one GB.  Each Riemann tensor's two antisymmetric index pairs must be assigned to different (of the two) epsilons in order for the Bianchi identity to function upon its variation, since $\Delta R_{abcd} \sim D_a D_c \Delta g_{bd}$. Which of the pairs' indices are contracted with those of the epsilons is irrelevant since the difference is at most one of sign, hence GB's uniqueness.  Consider $D=6$ for concreteness; there, counterterms and GB are of $6$-derivative order. The candidate one-loop invariants are then of three types,  
\begin{equation}
\int \hbox{Riem}^3, \, \, \, \, \, \, \, \, \, \, \, \int \hbox{Riem} \, d \, d\, \hbox{Riem}, \, \, \, \, \, \, \, \, \, \, \,  GB(6),
\end{equation}
($d$ denotes a covariant derivative), plus (harmless) terms containing at least one Ricci or $R$ factor. The single GB can only convert one of the two independent $\hbox{Riem}^3$ into the ``Ricci" type. The $\int \hbox{Riem} \, d\, d \hbox{Riem}$ can be reduced to $\hbox{Riem}^3$ form by the cyclic Bianchi identities $R_{\mu\nu [\alpha\beta;\rho]} =0$, $\hbox{div}\, \hbox{Riem} = \hbox{curl}\, \hbox{Ricci}$, to remove the $dd$ in favor of a possible extra $\hbox{Riem}$ where use of $[d, d] \sim \hbox{Riem}$ is needed. 
The same conclusions apply, {\it a fortiori}, to any higher $D$; a detailed list of curvature monomials for any $D$ can be found in [3].  A separate interesting set of actions, the Weyl invariant ones, also increases with dimension, beyond the unique $\int C^2$ of $4D$; a sample is given in [4], there being two $\int C^3$ (and one $\int C \Box C$) in 6D: there are now multiple Weyl ``gravities". 

\section{Comments}

We begin with a reminder, given the ambiguity in the literature, of its different uses of the words GB. The ambiguity is due to the appearance of terms resembling (1), the $D=10$ ``quadratic GB", in the slope expansion of strings, as discussed (correctly) long ago in [5]. Originally defined as the $D=4$ equivalent of the $D=2$ Euler invariant,
$\int d^2x , \eps^{ab} \, \eps^{cd} \, R_{abcd}$, (1) shares its property of being explicit metric-independent and a (purely global) number, its integrand being a total divergence. By abuse of language, the name GB has stuck to the quadratic in curvature combination in any number of dimensions, including odd ones, where the additional pairs of indices on the two $\eps$ tensors are saturated by explicit powers of the metric. These latter combinations are neither topological nor metric-independent, and indeed contribute to the field equations when added to the Einstein action $\int d^D R$ in any $D>4$. At linearized level, they would obviously not contribute to propagators, simply because their quadratic terms in terms of metric deviations are the same as they would be in $D=4$, namely zero, because the two curvatures exhaust the two relevant powers of $h_{\mu\nu}$. However, they would contribute as vertices since the cubic and higher terms do pick up the metrics' effects. Instead, the true GB terms as used here are of $R^n$ power and share the topological properties of Euler and $D=4$ GB; there is manifestly also only one such term per (even) $D$, by construction, that being one of our main points.  As stated earlier, being corrections, these effective string slope expansion additions to the Einstein action are not to to be treated as part of the original propagators or vertices around any vacuum.

We close with some conceivable, if unlikely, possible exceptions to these no-calculation statements. First, although there is no obvious selection principle that would allow only the one GB $(\hbox{Riem})^n$ counterterm (3) from among the many possible ones at higher dimensions, this should, and can, easily be checked explicitly by using the methods of [1]. Second, the catalog of [3] lists the locally independent monomials; there might be some integral relations among them -- that is, under the integral sign, they may differ from each other by divergences and terms involving Riccis. This too seems unlikely (except for the $\int \hbox{Riem}^p \Box^q \hbox{Riem}^{n-p-q}$ type above), simply because 4D has the special property that, to leading, $\sim h_{\mu\nu}^2$, order, all momenta are sandwiched between just two $h$, so on-Einstein shell where $0= \hbox{Ricci}=(pph)_{\mu\nu}$, there is only one $(h p^4 h)  \sim (\hbox{Riem})^2$ monomial. At higher $D$, the leading terms are now of the form $h pp \ldots h pp \ldots h \ldots$;  the $p$ and $h$ indices, are more ``segregated", which accounts for the increase of  $\int \hbox{Riem}^n$ combinations.

\section*{Acknowledgements.}
I thank J. Franklin for Latexterity.  This work was supported by grants NSF PHY-1266107 and DOE \# desc0011632.

\end{document}